\documentclass[12pt]{article}
\usepackage{apacite}
\usepackage{epsfig}
\usepackage{graphicx}
\usepackage{amssymb}
\usepackage{float}
\usepackage{subcaption}
\usepackage{caption}
\usepackage{mathtools}
\usepackage{natbib}

\title{Differentially Private Verification of Survey-Weighted Estimates}
\author{Tong Lin \thanks{South Hall 5521, University of California Santa Barbara, Santa Barbara, 93106, USA. E-mail: {\small \tt{tong\_lin@umail.ucsb.edu}}} \ and Jerome P. Reiter \thanks{Box 90251, Duke University, Durham, NC 27708, USA. E-mail: {\small \tt{jreiter@duke.edu}}}}
\date{}

\begin{document}

\maketitle

\begin{abstract}
    Several official statistics agencies release synthetic data as public use microdata files. In practice, synthetic data do not admit accurate results for every analysis. Thus, it is beneficial for agencies to provide users with feedback on the quality of their analyses of the synthetic data.  One approach is to couple synthetic data with a verification server that provides users with measures of the similarity of estimates computed with the synthetic and underlying confidential data. However, such measures leak information about the confidential records, so that agencies may wish to apply disclosure control methods to the released verification measures. We present a  verification measure that satisfies differential privacy and can be used when the underlying confidential are collected with a complex survey design. 
    We illustrate the verification measure using repeated sampling simulations where the confidential data are sampled with a probability proportional to size design, and the analyst estimates a population total or mean with the synthetic data.  The simulations suggest that the verification measures can provide useful information about the quality of synthetic data inferences.
\end{abstract}

\section{Introduction}\label{intro}

Survey sampling is widely used in various fields to make inferences about finite population quantities like population totals and averages. Typically, survey data are collected using complex sampling designs, such as stratified, probability proportional to size, or cluster sampling. These designs create unequal  probabilities that individuals will be selected into the sample.  Data analysts need to adjust for the unequal selection probabilities to obtain unbiased estimates of population quantities, for example, by using survey-weighted estimators.  

Many survey data sets are collected under pledges to protect the confidentiality of data subjects' identities and sensitive information.  As such, agencies seeking to disseminate survey data to the public typically apply some  redaction strategies to reduce the risks of unintended disclosures.  One approach is to generate synthetic data \citep{rubin:1993, little93, reiterpartsyn, drechsler, raghu,  reitertdp}, an approach taken for example, by the U.S.\ Bureau of the Census to share data from the Survey of Income and Program Participation.  In this approach, the agency simulates new values of confidential information using models estimated from the confidential data. These data are released as public use files, available for secondary data analysis. 

Naturally, the quality of inferences from synthetic data depend critically on the quality of the models used to generate the synthetic data \citep{reiter2005}. When the synthetic data models fail to accurately capture the distribution of the confidential data, secondary analysts of the synthetic data can obtain unreliable results. Thus, it is beneficial for agencies to provide means for secondary analysts to get feedback on the quality of their analyses of the synthetic data \citep{drechsler2stage}.

To do so, one approach is to provide secondary analysts access to a verification server \citep{vs09, mcclure2012towards, barrientos}. This is a computer system that has both the confidential and synthetic data.  The secondary analyst submits a query to the server for a measure of the similarity of estimates based on the confidential and synthetic data, for example, how far apart are the point estimates computed with the synthetic and confidential data.  The server reports back the verification measure to the analyst, who can decide if the synthetic data results are of adequate quality for their purposes. 

Verification measures leak information about the confidential data.  For example, \cite{vs09} illustrate how attackers could learn confidential information from targeted queries for verifications. Thus, it can be beneficial to apply disclosure treatment to verification measures  before release.  In particular, several researchers \citep[e.g., ][]{amitai, yureiter, barrientos} have developed verification measures that satisfy differential privacy \citep{dwork}. To date, however, researchers have not developed verification measures that satisfy differential privacy for survey-weighted analyses.  

In this article, we propose such measures.  The basic idea is to leverage the sub-sample and aggregate algorithm from the differential privacy literature \citep{nissim}.  We split the confidential data into disjoint subsets, estimate a survey-weighted analysis in each subset, determine the fraction  of these estimates falling within an analyst-specified distance of the synthetic data estimate, and add noise to this fraction using a Laplace Mechanism \citep{dworksmith}.  We investigate the performance of this approach using simulations of probability proportional to size sampling and a survey-weighted estimate of a population total or mean.  We consider settings where the synthetic data are representative of the population distribution and where they are not.  The simulation results suggest that the methods can provide useful feedback on the quality of synthetic data estimates of population totals  when the underlying confidential data are from a complex sample design.

The remainder of this article is organized as follows. In Section \ref{background}, we review survey-weighted estimates and differential privacy. In Section \ref{methods}, we describe our strategy for verification based on the sub-sample and aggregate algorithm.  We also discuss the settings of key parameters that affect the properties of the verification measures. In Section \ref{sims}, we use simulation experiments to illustrate the performance of the verification measures.  Finally, in Section \ref{conc}, we briefly summarize the main findings.

\section{Review of Survey-Weighted Estimation and Differential Privacy}\label{background}

In this section, we provide background useful for understanding the verification measures presented in Section \ref{methods}.

\subsection{Survey-weighted Estimation}

 Let $P$ be a finite population with $N$ elements, each with an index  $i = 1, \dots, N$. Let $X = (x_1, \dots, x_N)$ be the population values of some variable.  To motivate the methodology, we suppose that the analyst seeks inferences for the population total, $\tau = \sum_{i=1}^N x_i$.   We extend to estimation of population means in Section \ref{sec:sim:avg}.  Let $D$ be a subset of $P$ comprising $n$ elements randomly drawn from $P$.  We define the indicator $I_i=1$ if element $i$ is in the sample $D$, and $I_i=0$ otherwise. The vector $I = (I_1, \dots, I_N)$ represents the elements in $D$, and $n = \sum_{i=1}^N I_i$ is the sample size.

To determine $I$ and hence $D$, in this article we consider probability-proportional-to-size (PPS) sampling as an illustrative complex sampling design.  Let $Z=(z_1, \dots, z_N)$ be a numerical variable known for all $N$ units in $P$. We sample elements in $P$ with unequal probabilities proportional to $Z$. For each unit $i=1, \dots, N$, let $\pi_i = Pr(I_i =1)$ be its first-order inclusion probability. In PPS sampling of $n$ units, we have $\pi_i = nz_i/\sum_{i=1}^N z_i$. For any record $i$ where this quantity exceeds 1, we set that record's $\pi_i=1$. For the remaining records, we recompute the $\pi_i$ based on the sum of the $z_i$ in $P$ excluding the cases sampled with certainty.

For any probability sampling design including PPS sampling, a common approach to estimate $\tau$ is the  \cite{horvitz:thompson} estimator. We weight each sampled element by the inverse of its inclusion probability, and sum over all units in $D$.  More precisely, for $i=1, \dots, N$, let $w_i=1/\pi_i$.  We estimate $\tau$ using  
\begin{equation}
\hat{\tau} = \sum_{i \in D} x_i/\pi_i = \sum_{i \in D} w_i x_i.\label{ht}
\end{equation}
The estimator in \eqref{ht} is unbiased for $\tau$ for any sampling design, provided $\pi_i>0$ for  $i=1, \dots, N$.

\subsection{Differential Privacy}

Let $\mathcal{A}$ be an algorithm that takes a data set $D$ as input. We denote the output of $\mathcal{A}$ as $\mathcal{A}(D) = o$. We then define a neighboring data set $D^*$, which has the same data size as $D$. $D$ and $D^*$ differ in one row with all other rows identical. In accordance with the description provided by \citet{barrientos}, we present the definition of $\epsilon$-DP as follows.
\begin{itemize}
\item[] {\bf Definition 1 ($\epsilon$-differential privacy)}: An algorithm $\mathcal{A}$ satisfies $\epsilon$-differential privacy, abbreviated $\epsilon$-DP, if for any neighboring data sets $D$ and $D^*$, and any output $o \in Range(\mathcal{A})$, the  
\begin{equation}
Pr(\mathcal{A}(D) = o) \leq exp(\epsilon) Pr(\mathcal{A}(D^*) = o).\label{dpdef}
\end{equation}
\end{itemize}
 The $\epsilon$ is known as the privacy budget. It quantifies the similarity between the outputs of $\mathcal{A}$ being implemented over $D$ and $D^*$. Intuitively, smaller $\epsilon$ makes it more difficult for users to distinguish the data record that differs between $D$ and $D^*$, and thus guarantees a higher privacy level. 

DP has three important properties. Suppose that $\mathcal{A}_1$ and $\mathcal{A}_2$ are algorithms that satisfy $\epsilon_1$-DP and $\epsilon_2$-DP, respectively. First, for any data set $D$, releasing the outputs $\mathcal{A}_1(D)$ and $\mathcal{A}_2(D)$ satisfies $(\epsilon_1 + \epsilon_2)$-DP.  This is called the sequential composition property.  Second, for any two data sets $D$ and $E$ measured on disjoint sets of individuals, releasing the outputs of $\mathcal{A}_1(D)$ and $\mathcal{A}_2(E)$ guarantees max($\epsilon_1$, $\epsilon_2$)-DP. This is called the parallel composition property.  Third, for any algorithm $\mathcal{A}_3$, releasing the output $\mathcal{A}_3(\mathcal{A}_1(D))$ satisfies $\epsilon_1$-DP.  This is called the post-processing property.

One approach to achieve $\epsilon$-DP is the Laplace Mechanism \citep{dworksmith}.  Let $f$ be a function on $D \rightarrow \mathbb{R}^d$; for example, $f$ could sum the elements of one column (i.e., one variable) of $D$.  The global sensitivity is defined as 
$\Delta(f) = max_{(D, D^*)} \| f(D)-f(D^*) \|_1$
over all neighboring data sets $D$ and $D^*$. The Laplace Mechanism perturbs $f(D)$ by adding noise drawn from a Laplace distribution, i.e., we compute $f(D) + \eta$, where $\eta \sim Laplace(0, \, \Delta(f)/\epsilon)$. 

For some $f$, $\Delta(f)$ can be  large, resulting in a high probability of adding large noise to $f(D)$.  In such cases, we may want to satisfy $\epsilon$-DP using an  algorithm other than the Laplace Mechanism. 
One such mechanism, proposed by \citet{nissim}, is the sub-sample and aggregate algorithm. The basic idea is to randomly partition $D$ into $M$ disjoint subsets, $D' = \{D_1, \dots, D_M\}$. For each $D_k$, we determine $f(D_k)$ and then $f_{avg}(D') = \sum_{k=1}^M f(D_k)/M$. For many $f$, including our verification measures, changing $D$ by only one record changes the value of at most one $f(D_k)$. Thus, $\Delta(f_{avg})=\Delta(f)/M$.  We can apply a Laplace Mechanism to this $f_{avg}(D')$ using $\eta_{new} \sim Laplace(0, \Delta(f) / \epsilon M)$. Thus, we have reduced the variance of the noise significantly. We use the sub-sample and aggregate method to develop the differentially private, survey-weighted verification measures, as we now describe.

\section{Differentially Private, Survey-weighted Verification}\label{methods}

Suppose $D$ is a confidential data set comprising $i=1, \dots, n$ individuals measured on $p$ variables.  Thus, for any individual $i$, we have $D_i = (x_{i1}, \dots, x_{ip})$.  We also have a survey weight, $w_i = 1 / \pi_i$, where $\pi_i$ is the first-order inclusion probability of individual $i$. As a public use file, the agency generates a synthetic data set $D_0$ comprising $n_0$ simulated individuals with values of the same $p$ variables  in $D$. We assume that the agency generates $D_0$ following the approach in \citet{raghurubin}, in which it (i) simulates values for the $N-n$ records not in $D$ to create a completed population $P'$ and then (ii) takes a simple random sample of size $n_0$ from $P'$ that is released as $D_0$. Thus,  all $j=1, \dots, n_0$ synthetic individuals in $D_0$ have the simple random sample weights $N/n_0$.  The agency also might replace values for the records in $D$ when making $P'$; this does not affect our methodology.  For simplicity, we also assume that the agency releases only one synthetic data set, which is the case, for example, for the synthetic Longitudinal Business Database; see \citet{synlbd} and \citet{synlbd2}.  Our verification measures also can be applied when multiple implicates are released. We simply use the estimates from the synthetic data combining rules \citep{raghurubin} instead of the estimates from the one $D_0$.

 Suppose that the synthetic data analyst intends to estimate the population total of one of the variables, say $X$, based on $D_0$. For example, $x_i$ could be an indicator of whether individual $i$ speaks a certain language, so that $\tau = \sum_{i=1}^N x_i$ is the total number of people who speak that language in the population.  Let $\hat{\tau}_0= N \bar{x}_0 = N\sum_{j \in D_0} x_{j}/n_0$ be the synthetic data analyst's estimate of $\tau$ computed from $D_0$.  Let the synthetic data analyst's estimated variance of $\hat{\tau}_0$ computed with $D_0$ be  $\hat{\sigma}^2(\hat{\tau}_0) = N^2((1-n_0)/N)s^2_0/n_0,$ where $s^2_0 = \sum_{j \in D_0} (x_{j} - \bar{x}_0)^2/(n_0-1)$.

\subsection{Description of the Algorithm}\label{sec:descr}

To construct verification measures, we extend the approach  introduced by \citet{barrientos} and \citet{yang:reiter} to survey-weighted estimates.
 Let $\hat{\tau}$ be a survey-weighted estimate of $\tau$ computed with the confidential data $D$. The synthetic data analyst cannot compute $\hat{\tau}$, since they do not have access to $D$. However, we define it to motivate the verification algorithm. 

Let $\hat{d} = |\hat{\tau}_0-\hat{\tau}|$ be the absolute difference between $\hat{\tau}_0$ and $\hat{\tau}$. When $\hat{d}$ is small, where small is defined by the synthetic data analyst, it suggests that $\hat{\tau}_0$ is sufficiently accurate for the analyst's purposes.  We operationalize this by using a tolerance interval centered around $\hat{\tau}_0$, which we refer as $T(\hat{\tau}_0, \alpha)$. Here, $\alpha$ is a parameter that determines the width of the tolerance interval. To illustrate, suppose the synthetic data analyst views $D_0$ of adequate quality for their purposes if $\hat{\tau}_0$ is within three synthetic-data standard deviations of $\hat{\tau}$.  This analyst can set  $T=[\hat{\tau}_0 - 3\hat{\sigma}(\hat{\tau}_0), \, \hat{\tau}_0 + 3\hat{\sigma}(\hat{\tau}_0)].$  As another example, the analyst may decide that $\hat{\tau}_0$ is accurate enough as long as $\hat{\tau}$ is within some percentage of $\hat{\tau}_0$.  This analyst can set  $T(\hat{\tau}_0, \alpha) = [\hat{\tau}_0 \pm \alpha|\hat{\tau}_0|]$. The analyst then seeks to know whether $\hat{\tau} \in T(\hat{\tau}_0, \alpha)$.

To satisfy $\epsilon$-DP, however, the agency cannot directly release an indicator of whether $\hat{\tau} \in T(\hat{\tau}_0, \alpha)$. The agency should not use a Laplace Mechanism to perturb this indicator, as its global sensitivity equals one, making the Laplace distribution too high variance to return useful information. Further, generally it is not feasible to release a version of $\hat{\tau}$ that satisfies $\epsilon$-DP.  As noted in \citet{reiterarisa} and \citet{drechslersurvey}, to date there do not exist differentially private algorithms for releasing $\hat{\tau}$ from complex surveys that have low errors for reasonable privacy guarantees.

Instead, we use the sub-sample and aggregate method.  The verification server randomly partitions the confidential $D$ into $M$ disjoint subsets, with each partition denoted $D_k \in \{D_1, \dots, D_M\}$. The sample size of each $D_k$ is $n_k = \lfloor n/M \rfloor$. When $n$ is not  divisible by $M$, some partitions  have one more or one less unit than others. In each $D_k$, the server computes a survey-weighted population estimate of $\tau$ using only the data in $D_k.$  To do so, the server inflates each $w_i$ by a multiplicative factor of $n / n_k$. In particular, for $k = 1, \dots, M$, the server computes 

\begin{equation}
    \hat{\tau}_{k} = \sum_{i \in D_k} w_i(n/n_k)x_i.
\end{equation}

In each $D_k$, the synthetic data analyst specifies a tolerance interval $C(\hat{\tau}_0, \alpha, \gamma)$. This  interval is not necessarily the same as $T(\hat{\tau}_0, \alpha)$. We discuss ways of setting $C(\hat{\tau}_0, \alpha, \gamma)$ in Section \ref{sec:tolerance}.  For $k=1, \dots, M$, let  $A_k=1$ when $\hat{\tau}_{k} \in C(\hat{\tau}_0, \alpha, \gamma)$, and $A_k=0$ otherwise. Let $S = \sum_{k=1}^M A_k$ be the number of partitions where $A_k=1$. 
Then, $S/M$ is an estimate of the probability that, for an arbitrary $D_k$, the $\hat{\tau}_{k} \in C(\hat{\tau}_0, \alpha, \gamma)$. Values of $S/M$ near 1 indicate that the confidential-data estimates in the partitions frequently fall inside the tolerance intervals, which suggests that the estimates from the confidential data are similar to the estimate from the synthetic data. Values of $S/M$ near 0 indicate that estimates from confidential data are dissimilar to the estimate from the synthetic data, suggesting the synthetic data estimate is not sufficiently accurate for the analyst's purposes.

To meet the $\epsilon$-DP requirement, the verification server needs to add noise to $S$. We do so via the Laplace Mechanism. The server randomly draws a sample $\eta \sim Laplace(0, 1/\epsilon)$  and sets $S^R=S+\eta$. This Laplace Mechanism presumes a $\Delta(f) = 1$, that is, changing one record in $D$  only affects at most one $A_k$.  At the end of this section, we discuss the privacy properties of $S^R$ in more detail.

Because $S^R/M$ can be outside $[0,1]$, we apply post-processing to enhance the interpretability of the reported verification measure. Specifically, we assume that each $A_k \sim Bernoulli(r)$, where $r$ is the probability that any randomly generated $\hat{\tau}_{k} \in C(\hat{\tau}_0, \alpha, \gamma)$. Hence, we assume that $S|r \sim Binomial(M,r)$. We suppose a uniform prior distribution for $r$, which equivalently is $r \sim Beta(1,1)$ where $Beta$ represents a Beta distribution. Thus, the model for post-processing $S^R$ is 
\begin{equation}
S^R|S \sim Laplace(S, 1/\epsilon) \,\,\,\,\,\, 
S|r \sim Binomial(M, r) \,\,\,\,\,\,\, r \sim Beta(1,1).
\end{equation}

We obtain the posterior distribution $p(r|S^R)$
via a Gibbs sampler.  The sampler does not use the true value of $S$, which is unavailable to the algorithm to maintain $\epsilon$-DP.  Rather, we average over plausible values of $S$.
The full conditional for $r$ is
\begin{align}
    p(r|S, \,S^R) 
    & \propto Pr(S|r)Pr(r) \propto r^S(1-r)^{M-S}
\end{align}
which is the kernel of a $Beta(S+1,\, M-S+1)$ distribution.
The full conditional for $S$ is 
\begin{align}
    Pr(S|r, \, S^R) 
    & \propto Pr(S^R|S)Pr(S|r)  \propto e^{-\frac{|S^R-S|}{1/\epsilon}} \frac{1}{\Gamma(S+1)\Gamma(M-S+1)}r^S(1-r)^{M-S}.
\end{align} 
The verification server releases draws from $p(r |S^R)$, including the posterior median. 

When $p(r |S^R)$ is concentrated near 1, the analyst can conclude that the synthetic and confidential data offer similar estimates of $\tau$.  When $p(r |S^R)$ is concentrated near 0, the analyst can conclude that the synthetic and confidential data estimates of $\tau$ are too dissimilar for $\hat{\tau}_0$ to be considered sufficiently accurate.  Values of $r$ near 0.5 suggest that the evidence is unclear.

Finally, we close this section with a discussion of the privacy protection properties of these verification measures.  First, because of the post-processing property of $\epsilon$-DP mentioned in Section \ref{background}, releasing $p(r|S^R)$ does not affect the privacy guarantee endowed by generating $S^R$.  The Bayesian modeling only uses $S^R$; it never uses other results from the confidential data.  Second, presuming $\Delta(f)=1$ for the verification measures implicitly presumes that changing one individual in $D$ does not change the data, including the survey weights, for any other individuals in $D$. This could be violated, for example, when the agency adjusts survey weights for nonresponse or does data editing by using information from multiple records. We leave accounting for this possibility to future research.  Third, we note that $D_0$ may not satisfy $\epsilon$-DP; indeed, most implementations of synthetic data to date do not.  As a result, we cannot rely on the sequential composition property of $\epsilon$-DP to quantify the privacy loss from releasing both $S^R$ (or $p(r|S^R)$) and $D_0$.  Of course, if $D_0$ (or more precisely $\hat{\tau}_0$) is differentially private \citep[as in, e.g., ][]{bowen, fangliu}, then the sequential composition property applies.  Thus, agencies and analysts can interpret the privacy protection afforded by the verification measures as a bound on the additional privacy leakage due to releasing the verification measure over the leakage from releasing $D_0$ itself.

\subsection{Specifying the Tolerance Interval} \label{sec:tolerance}

The synthetic data analyst needs to specify  $C(\hat{\tau}_0, \alpha, \gamma)$. Here, $\gamma$ plays the role of an inflation factor that may be used to go from $T(\hat{\tau}_0, \alpha)$, which is based on a sample size of $n$, to $C(\hat{\tau}_0, \alpha, \gamma)$, which is based on  a sample size of approximately $n/M$. Following \citet{yang:reiter}, we consider two classes of tolerance intervals.  First, the analyst may set $C(\hat{\tau}_0, \alpha, \gamma)=T(\hat{\tau}_0, \alpha)$; we call this a fixed tolerance interval.  
To illustrate, suppose $\hat{\tau}_0 = 100000$ and $\hat{\sigma}(\hat{\tau}_0) = 1000$. The analyst wants to know if $\hat{\tau}$ falls within $10\%$ of $\hat{\tau}_0$, i.e., within $10000$. For a fixed tolerance interval, we have $T(\hat{\tau}_0, \alpha) = C(\hat{\tau}_0, \alpha, \gamma) = [90000, \, 110000]$. 

Alternatively, the analyst may set  $C(\hat{\tau}_0, \alpha, \gamma) \neq T(\hat{\tau}_0, \alpha)$; we call this an adjusted tolerance interval.
The main motivation for adjusted tolerance intervals is that the smaller sample size in any $D_k$ increases the variance associated with $\hat{\tau}_k$ compared to the variance of $\hat{\tau}$ from $D$.  If we use a fixed tolerance interval with $C(\hat{\tau}_0, \alpha, \gamma) = T(\hat{\tau}_0, \alpha)$, any $\hat{\tau}_k$ has increased probability of falling outside $C(\hat{\tau}_0, \alpha, \gamma)$ even when $\hat{\tau} \in T(\hat{\tau}_0, \alpha)$. Thus, we use the parameter $\gamma$ to inflate the tolerance intervals within the partitions.

To do so, we follow the strategy used by  \citet{barrientosjcgs}, which we explain using an illustrative example.  
Suppose the analyst has in mind $T(\hat{\tau}_0, \alpha) = [\hat{\tau}_0 \pm 3\hat{\sigma}(\hat{\tau}_0)]$. Here, we set $\alpha=3$, although analysts could choose other values, e.g., $\alpha=10$ for a tolerance of $\pm 10000$ when $\hat{\sigma}(\hat{\tau}_0) = 1000$. Suppose we have $M=25$ disjoint partitions, $(D_1, \dots, D_{25})$.  In this case, it can be reasonable to  approximate $\hat{\sigma}(\hat{\tau}_{k})$ with $\sqrt{n/n_k}\hat{\sigma}(\hat{\tau})$, that is, we inflate the variance to recognize the change in sample size going from $D$ to $D_k$. We use the inflated standard error when constructing the adjusted tolerance interval, so that  
$C(\hat{\tau}_0, \alpha, \gamma) = [\hat{\tau}_0 \pm (5)3 \cdot \hat{\sigma}(\hat{\tau}_0)]$. Here, $\gamma = \sqrt{25}=5$. As a default, we recommend setting $\gamma = \sqrt{M}$ for adjusted intervals.  We note that $\gamma = 1$ in the fixed tolerance intervals.

\subsection{Choosing $M$}\label{sec:chooseM}

In this section, we discuss the choice of the number of partitions $M$. We consider the effect of changing $M$ on $S/M$ itself and on the noise from the Laplace Mechanism.  This discussion closely follows that in \citet{yang:reiter}. 

By design, $S$ and hence $S/M$ can be one of  $M+1$ values. For instance, when $M = 5$, we have $S/M \in \{0, 0.2, 0.4, \dots, 1\}$. In this case, $S/M$ might not be granular enough for the analyst to make clear interpretations of the quality of $D_0$. In addition, with a small $M$, the perturbation from the Laplace Mechanism will have a greater proportional impact on $S$, potentially making it more difficult to interpret $S^R$. On the other hand, for a given $D$, fewer partitions means larger sample sizes in  each $D_k$. Larger values of $n_k$ reduce the variance of $\hat{\tau}_k$ in each partition, which can result in more reliable inferences about the differences in $\hat{\tau}$ and $\hat{\tau}_0$. Finally, a small $M$ can increase  the variance of $S/M$ over the random partitions. 

Analysts need to balance these trade offs in selecting $M$. Overall, the goal is to choose an $M$ so that the verification results are consistent with the results that could be obtained using the full confidential data.  In other words, we want $Pr(\hat{\tau} \in T(\hat{\tau}_0, \alpha))$ to be close to $Pr(\hat{\tau}_{k} \in C(\hat{\tau}_0, \alpha, \gamma))$. Specifically, if $\hat{\tau} \in T(\hat{\tau}_0, \alpha)$, the probability density of $S^R/M$ should have most mass near 1. When $\hat{\tau}$ is outside $T(\hat{\tau}_0, \alpha)$, the probability density of $S^R/M$ should have most mass near 0. In Section \ref{sims}, we present simulation studies with different $M$ to help inform this decision.

\section{Simulation Studies}\label{sims}

In this section, we conduct simulation studies to illustrate the properties of the verification measures. 
We first generate a population $P$ comprising $N=10000000$ individuals. For each individual $i$, we generate two variables $(z_i, x_i)$ sampled from $z_i \sim Uniform(0, 10)$ and $x_i|z_i \sim N(z_i+5, 2).$  For each unit $i$ in $P$, we assign an inclusion probability proportional to $z_i$, so that $\pi_i = n z_i/ \sum_{i=1}^N z_i$ where $n$ is the sample size. Using $\pi_i$, we take a PPS sample from $P$ to make the confidential data $D$. 

We generate synthetic data from $D$ using two strategies. 
The first method involves generating a $D_0$ that is representative of $P$. To do so, we need to account for the complex design when synthesizing. Failure to do so can result in synthetic data that do not look like $P$. Since our goal is to evaluate the verification measures rather than implement a synthesizer that handles survey weights \citep[e.g., ][]{kim:drechsler, hu:savitsky}, we simply take a simple random sample of size $n_0$ from $P$ to create $D_0$. Of course, this is not possible in genuine applications; agencies need to account for the complex design using $D$ to make $D_0$.  However, given that we know $D_0$ is an accurate representation of $P$, ideally the verification measures indicate that the synthetic data provide accurate estimates. 

In the second method, we generate $D_0$ directly from $D$ but ignore the sampling design. Specifically, we randomly draw $n_0$ samples from $\mathcal{N}(\bar{x}, s^2_x)$, where $\bar{x}$ and $s^2_x$ are the sample mean and variance of the variable $X$ in $D$. This synthesizer should lead to inaccurate estimates since $D_0$ is not representative of $P$.  Thus, it allows us to examine the performance of the verification measure when $D_0$ offers unreliable estimates.

We focus on factors that could affect the performance of the verification algorithm, namely $M$, $n_k$, and the  tolerance intervals. We consider $n_k \in \{500, 20000, 50000\}$ and $M \in \{25, 50, 90\}$ partitions. For each combination of $n_k$ and $M$, we draw $n=n_k M$ samples from $P$ using PPS sampling to  make $D$. We set $n_0 = n$. We repeat the steps for generating $(D,D_0)$ for 200 times for each of the two synthetic generation methods.   We set $\epsilon=1$ for all measures.  For  the tolerance intervals, we use  a fixed tolerance interval of 
$T(\hat{\tau}_0, \alpha) = [\hat{\tau}_0 - \alpha \hat{\sigma}(\hat{\tau}_0), \hat{\tau}_0 + \alpha \hat{\sigma}(\hat{\tau}_0)]$.   For the adjusted interval, we set $\gamma = \sqrt{M}$ and $C(\hat{\tau}_0, \alpha, \gamma) = [\hat{\tau}_0 - \alpha \sqrt{M} \hat{\sigma}(\hat{\tau}_0), \, \hat{\tau}_0 + \alpha \sqrt{M} \hat{\sigma}(\hat{\tau}_0)]$. We consider $\alpha \in \{1,\,3, \,5 \}$.

For each pair of $D$ and $D_0$, we compute two quantities. First, we define a binary variable $Q$, which is an indicator that takes value of 1 when $\hat{\tau}$ is inside the original tolerance interval, i.e., $Q = \mathbb{I}(\hat{\tau} \in T(\hat{\tau}_0, \alpha))$. For the 200 pairs of $(D_0,D)$, we get $Q_1, \dots, Q_{200}$. We then calculate $r_{full} = \sum_{i=1}^{200} Q_i/200$, which is an approximate estimate of $Pr(\hat{\tau} \in T(\hat{\tau}_0, \alpha))$. Of course, the synthetic data analyst does not get $Q$ or $r_{full}$, as they have only the differentially private results.  Nonetheless, we can use $r_{full}$ to evaluate the differentially private measures.  Second, with each $(D, D_0)$, we implement the differentially private verification measure to compute the posterior distribution of $r$. We store the posterior medians of $r$.  Ideally, within any simulation setting, the posterior medians of $r$ are similar to $r_{full}$, indicating that the differentially private verification measure tends to result in similar conclusions as using the original interval.

\subsection{Results for Synthesis Based on SRS of $P$}  \label{sec:sim:srs}

\begin{figure}[t]
    \centering  
    \includegraphics[width=5.5in]{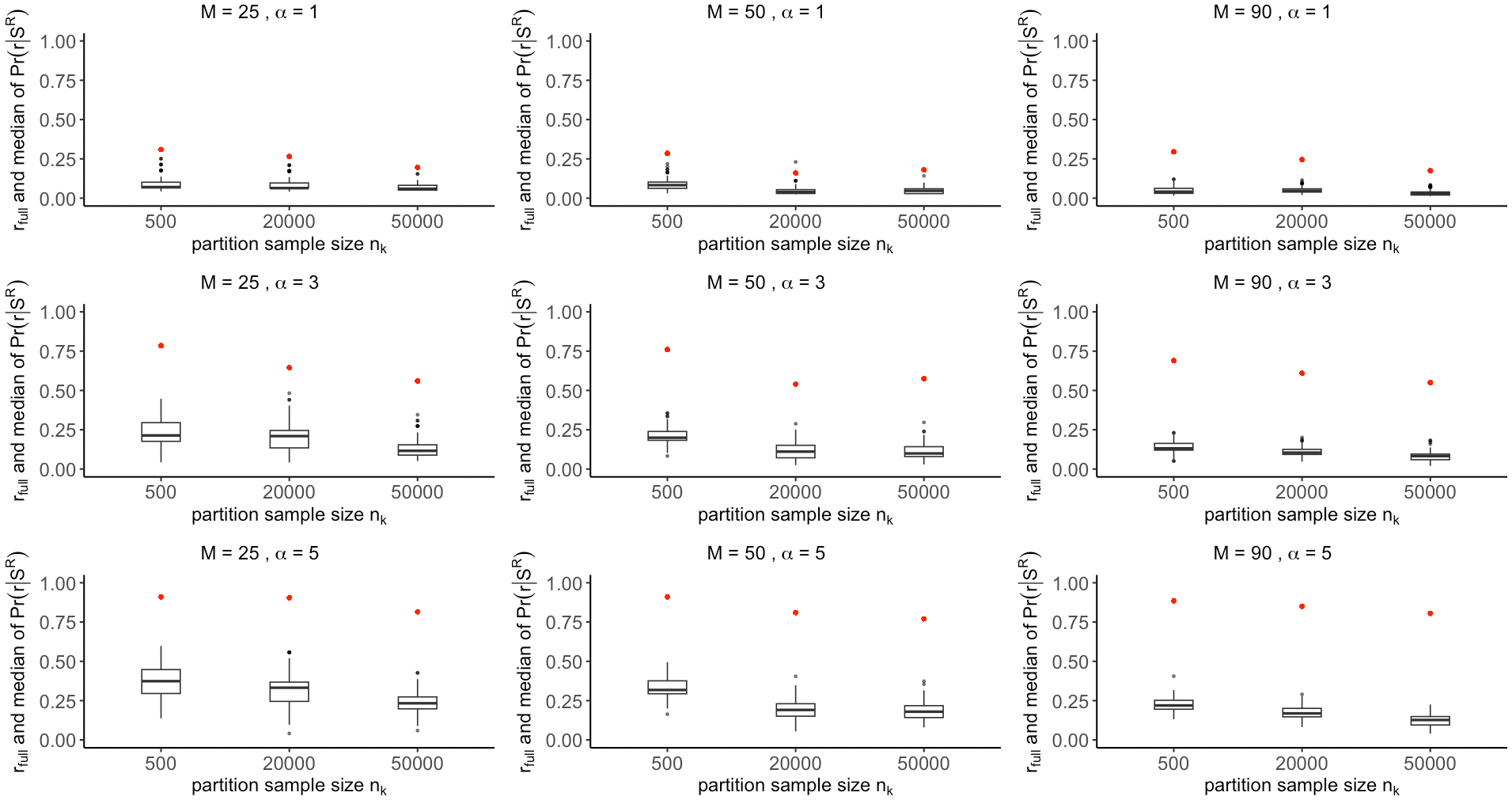}
    \caption{$r_{full}$ (red points) and posterior medians of $r$ (box plots) using fixed tolerance intervals for the population total. Synthetic data are a SRS from $P$.}
    \label{fig:fix1}
\end{figure}

 Figure \ref{fig:fix1} summarizes the results for the fixed tolerance intervals when the synthesizer faithfully represents $P$.  There is obvious discrepancy between the values of $r_{full}$ and posterior medians of $r$, which indicates inconsistency between the conclusions drawn from using the full data set and the partitions. The posterior medians of $r$ are always much smaller than their corresponding $r_{full}$, for all $\alpha$ considered.  Quite simply, the verification measure with a fixed tolerance interval does not have acceptable performance.

\begin{figure}[h]
    \centering  
    \includegraphics[width=5.5in]{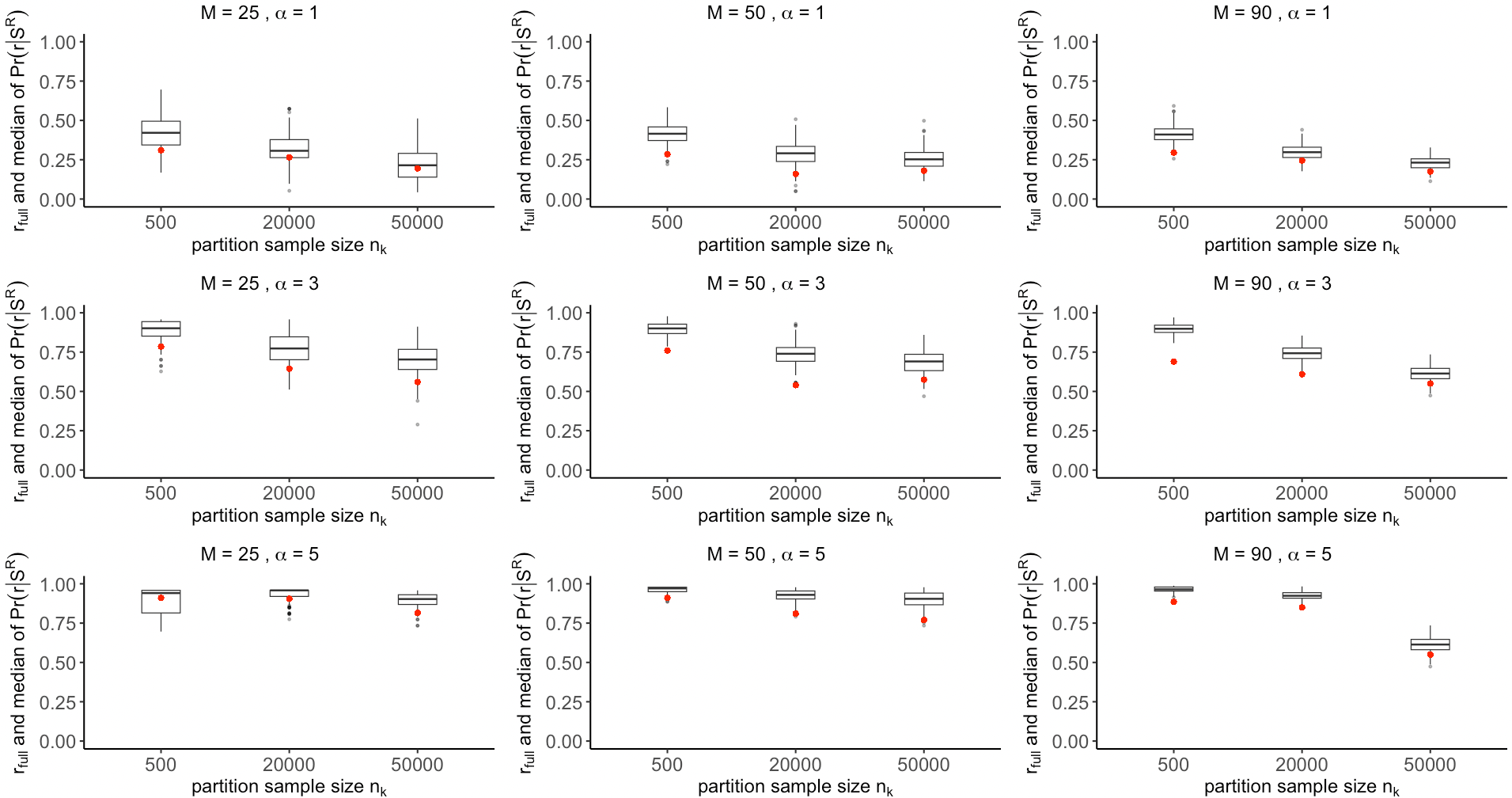}
    \caption{$r_{full}$ (red points) and posterior medians of $r$ (box plots) using adjusted tolerance intervals  for the population total. Synthetic data are a SRS from $P$.}
    \label{fig:vary1}
\end{figure}

We next turn to the results for the adjusted tolerance interval, displayed in Figure \ref{fig:vary1}. 
In most instances, the posterior medians of $r$ are close enough to the values of $r_{full}$ that analysts likely would reach similar conclusions about the verification using either $r$ or $r_{full}$. 
When $\alpha = 1$,  $r_{full}$ and the posterior medians of $r$ are typically around 0.3. When $\alpha = 3$, the value of $r_{full}$ increases to between 0.5 and 0.75. The majority of the posterior medians of $r$ tend to be larger than $r_{full}$, suggesting some over-optimism in the verification decision. When $\alpha = 5$, $r_{full}$ and the posterior medians of $r$ tend to be above 0.8. 

Holding constant $M$ and $\alpha$, we see that smaller values of $n_k$ correspond to larger values of both $r_{full}$ and medians of $r$. Evidently, in these simulations, decreasing $n_k$ increases the probability that $\hat{\tau}$ and $\hat{\tau}_{k}$ are within the analyst's tolerance interval. However, $r_{full}$ and the posterior medians of $r$ tend to track each other for all $n_k$ considered. Of course, the trend

\clearpage

\noindent may not hold if $n_k$ gets very small, as the variance of $\hat{\tau}_k$ may become so large as to make $S/
M$ go toward zero, particularly when the tolerance interval is tight around $\hat{\tau}_0$ compared to the variance of $\hat{\tau}_k$.

Holding constant $n_k$ and $\alpha$, the changes in $M$ have little effect on the average value of the posterior medians of $r$ in this simulation. Nonetheless, the variance of the posterior medians of $r$ decreases as $M$ grows larger. This is expected: increasing $M$ reduces the impact of the noise from the Laplace Mechanism on $S/M$, and thus reduces the variance in $S^R/M$.  As a default, we recommend setting $M=20$ or $M=25$ to ensure a fine enough grid while ideally keeping reasonably large sample sizes within the partitions.

\subsection{Results for Biased Synthesis}\label{sec:sim:bad}

\begin{figure}[t]
    \centering  
    \includegraphics[width=5.5in]{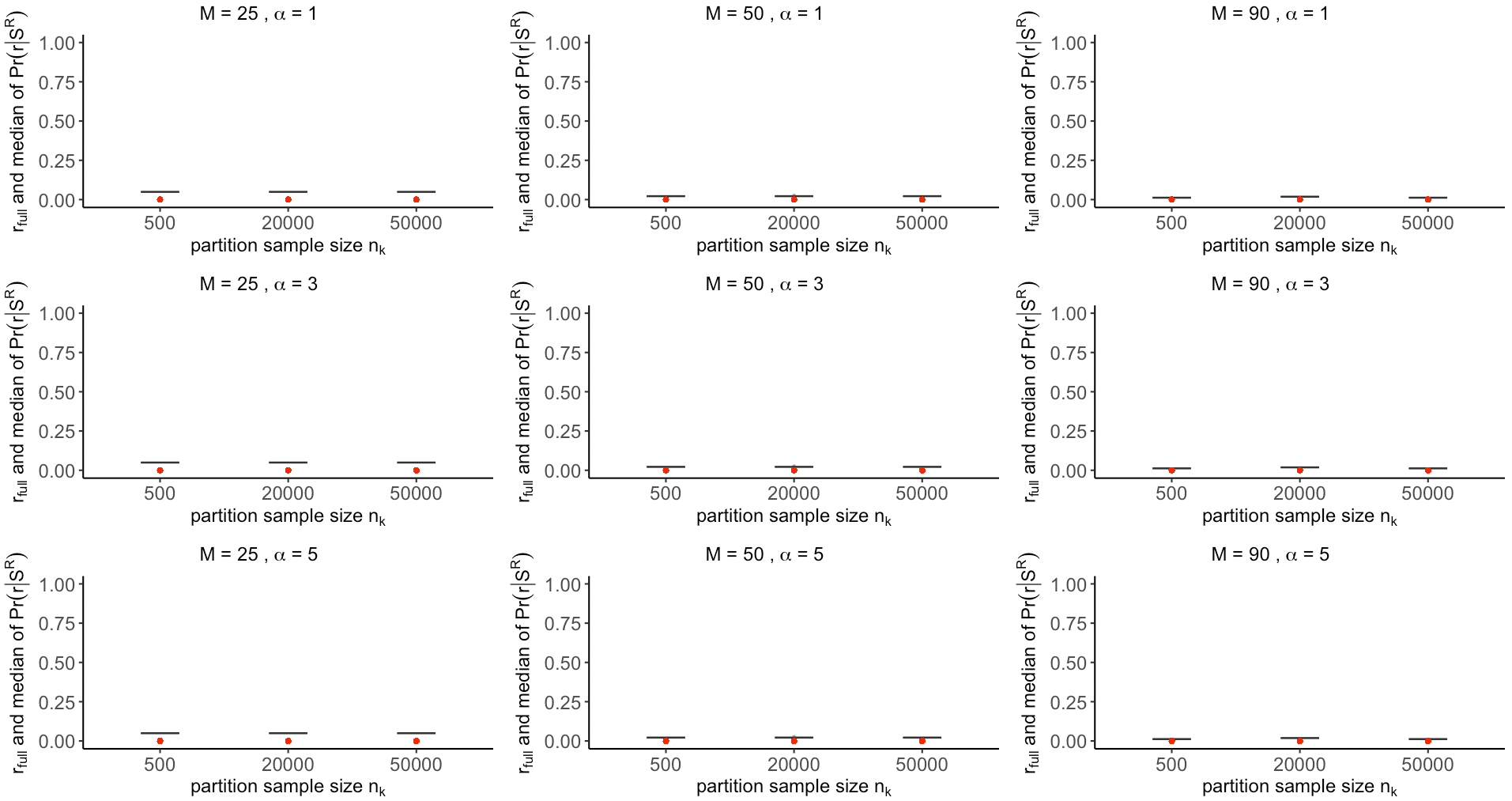}
    \caption{$r_{full}$ (red points) and posterior medians of $r$ (box plots) using adjusted tolerance intervals for the population total. Synthetic data are a biased sample.}
    \label{fig:vary_unwt}
\end{figure}

We now turn to the results from the simulation where the agency disregards the sampling design when generating $D_0$. We expect these synthetic data to be low quality for estimating $\tau$ and desire the verification measures to indicate as such. 

We first provide evidence that accounting for the survey design is important in this simulation. 
For each generated $D$, we estimate $\tau$ using both the \citet{horvitz:thompson} estimator and an unweighted estimator $N\bar{x}$. The true value is $\tau= 99984562.$
While the \citet{horvitz:thompson} estimator is unbiased, the averages of $N\bar{x}$ across the simulation settings tend to be around 117000000, which is much larger than $\tau$.

We implement the verification procedure using the biased synthetic data. Because the fixed tolerance intervals performed poorly in Section \ref{sec:sim:srs}, we only display the results for the adjusted tolerance intervals, shown in Figure \ref{fig:vary_unwt}. 
Regardless of the value we set for $n_k$, $M$, and $\alpha$, $r_{full}$ and the posterior medians of $r$ are close to 0. The poorly generated synthetic data lead to a  biased estimate of $\tau$, so that the $\hat{\tau}_k$ tend not to lie within the tolerance interval. Evidently, the verification measures appropriately clue the analyst that the synthetic data are unreliable for estimating $\tau$ accurately.

\subsection{Simulation Studies with a Population Average}\label{sec:sim:avg}

As an additional set of studies, we repeat the simulations from Section \ref{sec:sim:srs} and \ref{sec:sim:bad} using the population average, $\bar{X}= \sum_{i=1}^Nx_i/N$.  For the synthetic data, the estimate is simply $\bar{x}_0 = \sum_{j\in D_0} x_j/n_0$, with estimated variance $\hat{\sigma}_0 = (1-n_0/N)s^2_0/n_0$.
For the confidential data $D$ and each partition $D_k$,  we estimate $\bar{X}$ using survey-weighted ratio estimators.  Re-using $\hat{\tau}$ and $\hat{\tau}_k$ for convenience, we have 
\begin{eqnarray}\label{eq:ppsmean}
   \hat{\tau} &=& \sum_{i \in D} w_ix_i /\sum_{i \in D} w_i \\
    \hat{\tau}_{k} &=& \sum_{i \in D_k} w_i (n/n_k) x_i/\sum_{i \in D_k} w_i (n/n_k).
\end{eqnarray}
These are the usual estimators of $\bar{X}$ for PPS samples as well as other common designs. 

\begin{figure}[t]
    \centering  
    \includegraphics[width =5.5in]{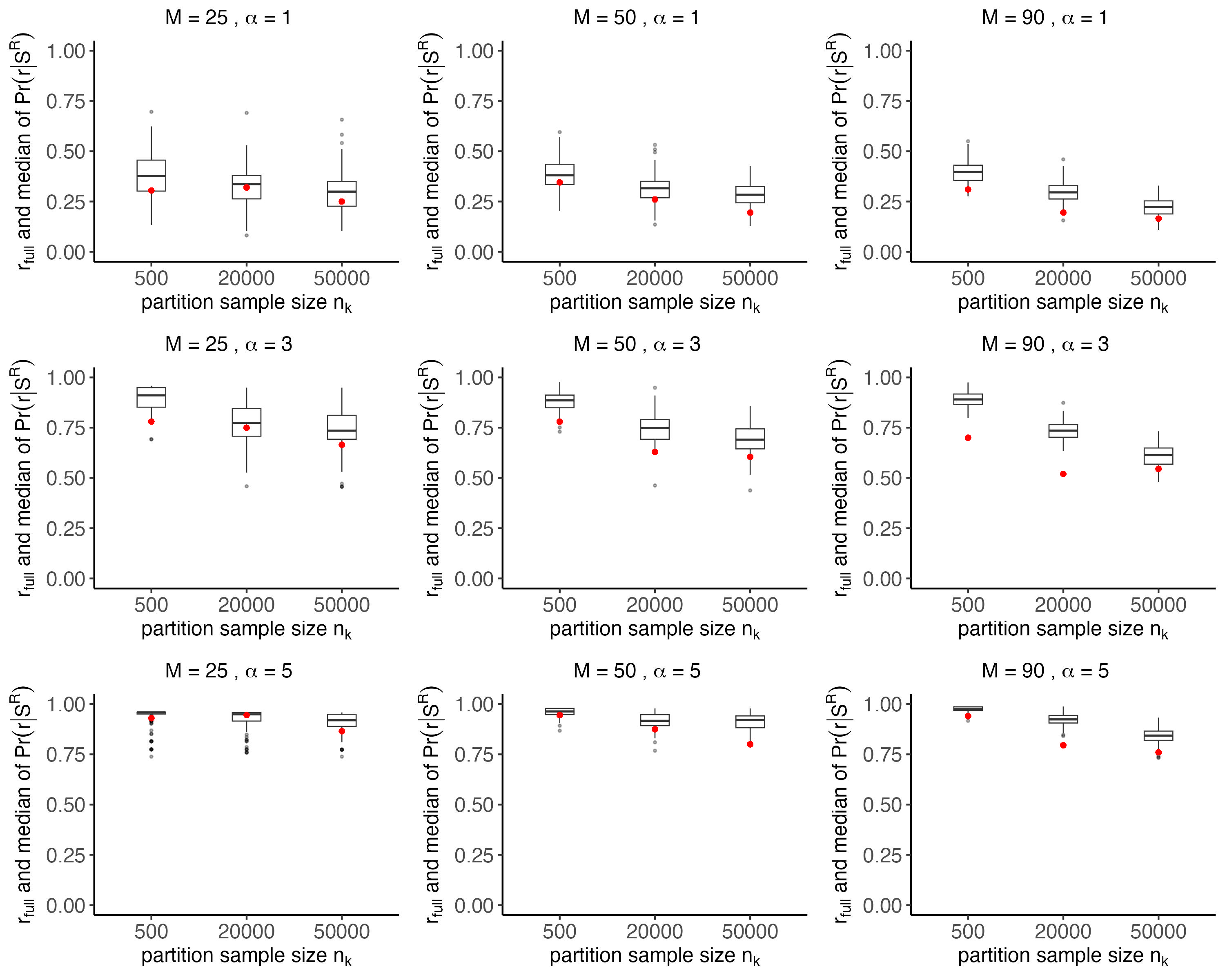}
    \caption{$r_{full}$ (red points) and posterior medians of $r$ (box plots) using adjusted tolerance intervals for the population average. Synthetic data are a SRS from $P$.}
    \label{fig:vary1_avg}
\end{figure}

\begin{figure}[h]
    \centering  
    \includegraphics[width=5.5in]{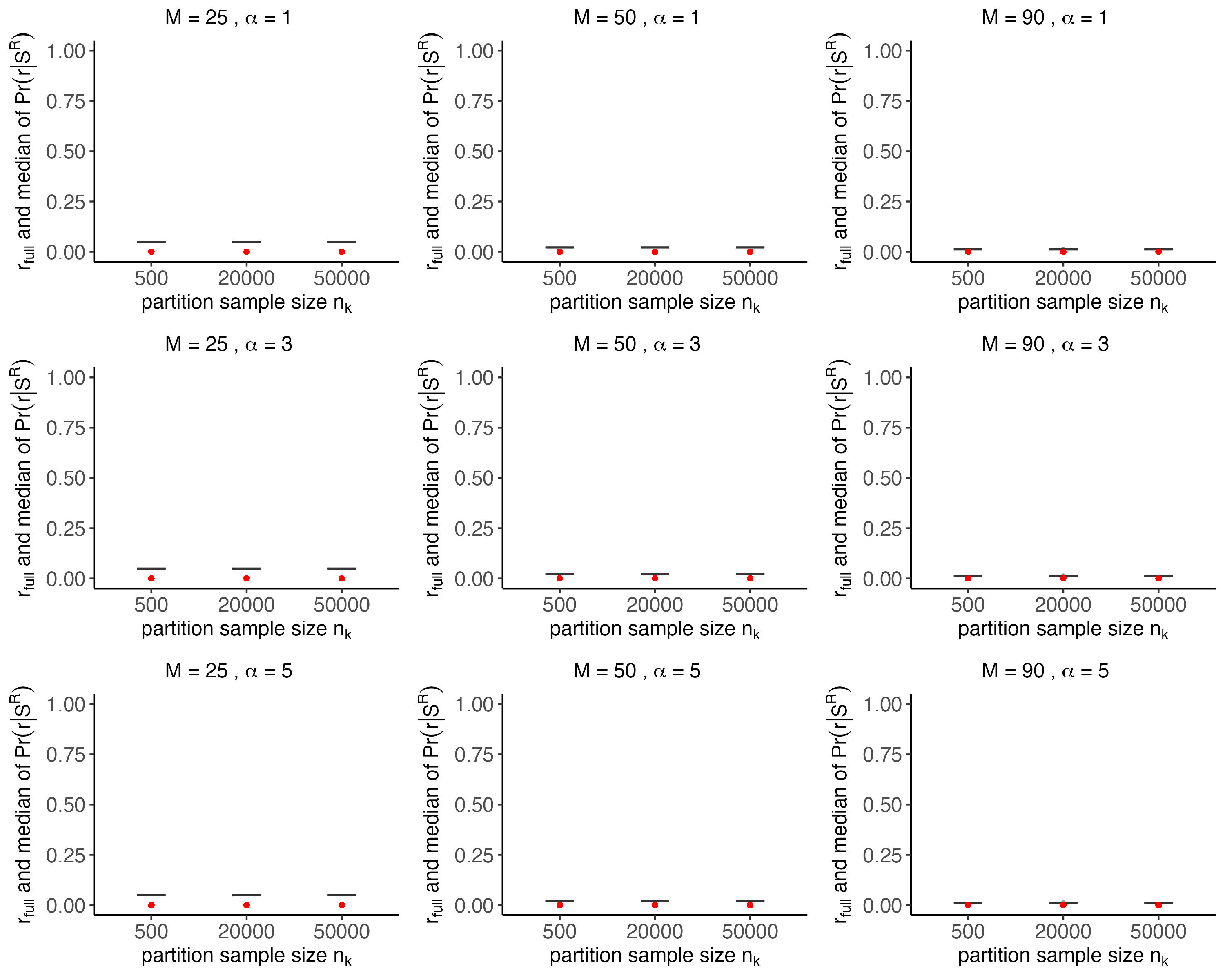}
    \caption{$r_{full}$ (red points) and posterior medians of $r$ (box plots) using adjusted tolerance intervals for the population average. Synthetic data are a biased sample.}
    \label{fig:vary_avg_unwt}
\end{figure}

The patterns in the simulations using the population average mimic those for the population total.  In particular, the fixed tolerance interval does not perform well, displaying properties similar to those in Figure \ref{fig:fix1}; we do not display these results here.  The adjusted tolerance interval performs reasonably well, especially when $M=25$, as evident in Figure \ref{fig:vary1_avg} for the design where the synthetic data come from a good-fitting model and in Figure \ref{fig:vary_avg_unwt} when the synthetic data come from a biased model.  Overall, the results suggest the verification measures can be useful for population averages as well as totals.

\section{Final Remarks}\label{conc}

In this article, we address the gap in existing verification measures for synthetic data when the underlying confidential data come from a complex survey design. 
Our findings in the simulation experiments suggest that adjusted tolerance intervals tend to yield more reliable verifications than fixed tolerance intervals. Hence, we recommend using adjusted tolerance intervals as a general practice.  Of course, as with all simulation studies, these findings are specific to the simulation design presented here.  We recommend that agencies and analysts  undertake their own simulation studies to assess the properties of the verification measures for their settings.  Ideally, such studies can be based on artificial data, like those presented in Section \ref{sims}, tuned to match the sampling design and population characteristics, so that agencies need not use additional privacy budget for simulations.

\clearpage

\bibliographystyle{natbib}
\bibliography{lit}

\end{document}